\documentclass[aps,twocolumn,nofootinbib,nobibnotes,superscriptaddress,preprintnumbers,longbibliography]{revtex4-1}

\usepackage[dvips]{graphicx}
\usepackage[english]{babel}
\usepackage[normalem]{ulem}
\usepackage[utf8]{inputenc}
\usepackage[usenames,dvipsnames]{xcolor}
\usepackage[colorlinks=true,pdfstartview=FitV,linkcolor=blue,citecolor=blue,urlcolor=blue]{hyperref}
\usepackage{amsfonts,amsmath,amssymb,amsthm,mathrsfs,mathtools}
\usepackage{comment,enumerate,footnote,graphicx,relsize,subfloat}
\usepackage{afterpage,array,framed,multirow,tabu,tabularx}
\usepackage{epsfig,color,cancel}
\usepackage{bm}
\usepackage{cleveref}
\usepackage{graphics}
\usepackage{slashed}
\usepackage{soul}
\usepackage{verbatim}

\def\mdm		{m_{\mathsmaller{\rm DM}}}

\newcommand{\be}{\begin{equation}}
\newcommand{\ee}{\end{equation}}
\crefformat{pluralequation}{#2\black{eqs.~(}#1\black{)}#3}
\Crefformat{pluralequation}{#2\black{Equations~(}#1\black{)}#3}
\crefformat{pluralfigure}{#2\black{figs.~}#1#3}
\Crefformat{pluralfigure}{#2\black{Figures~}#1#3}

\begin{document}

\title{
Acoustically driven dark matter freeze-out
}

\author{Iason Baldes}
\affiliation{Laboratoire de Physique de l'\'Ecole Normale Sup\'erieure, ENS, \\ Universit\'e PSL, CNRS, Sorbonne Universit\'e, Universit\'e Paris Cit\'e, F-75005 Paris, France}

\author{Ryusuke Jinno}
\affiliation{Department of Physics, Graduate School of Science, Kobe University, 1-1 Rokkodai, Kobe, Hyogo 657-8501, Japan}

\begin{abstract}
We extend the study of the effect of density perturbations to the well known thermal dark matter freeze-out scenario. We find $\sim 10 \, \%$ enhancements in the cross section are required to match onto the observed relic abundance for primordial curvature perturbations $\sim 0.2$ at length scales $\sim 1/$Hubble at freeze-out. Such corrections may be of importance in scenarios in which such perturbations are present and observational signals, such as the indirect detection rate, depend sensitively on the DM mass and freeze-out cross section, e.g.~near resonances associated with DM bound states.
\end{abstract}

\preprint{KOBE-COSMO-25-10}

\maketitle

\section{Introduction}
\label{sec:introduction}

The freeze-out of particles in the presence of density perturbations, in a radiation dominated universe, so-called ``acoustically driven freeze-out," has been studied in the cases of type-I~\cite{Hotokezaka:2025ewq}, type-II and III leptogenesis~\cite{Baldes:2025uwz}. For some related work on big bang nucleosynthesis (BBN), see~\cite{Inomata:2016uip}.
Here we extend the study of this effect to freeze-out of dark matter (DM), considering large amplitude perturbations at small scales during freeze-out. These may arise either from inflation~\cite{Mukhanov:1981xt}, preheating~\cite{Kofman:1994rk,Kofman:1997yn}, topological defects~\cite{Hindmarsh:1994re,Vilenkin:2000jqa,Durrer:2001cg}, or a strong early universe phase transition~\cite{Liu:2022lvz,Elor:2023xbz,Lewicki:2024ghw,Zou:2025sow,Franciolini:2025ztf}, which may itself be associated with the physics of the DM sector~\cite{Hambye:2013dgv,Baldes:2017rcu,Hambye:2018qjv,Baldes:2018emh,Baldes:2020kam,Baldes:2021aph,Kierkla:2022odc,Wong:2023qon,Balan:2025uke}. 

The effect of large scale perturbations on freeze-out and freeze-in DM has previously been studied in Ref.~\cite{Holst:2023msh}. Such large scale perturbations are constrained to be of much smaller amplitude by CMB observations, than the small scale perturbations we will consider in this paper. Along the same lines, both large and small scale perturbations for freeze-in were examined in Ref.~\cite{Stebbins:2023wak}. Moreover, the effect of perturbations for the relic density of axions has recently been determined in Refs.~\cite{Eroncel:2025qlk,Bodas:2025eca}.

Acoustically driven freeze-out will necessarily lead to isocurvature perturbations, at sub-horizon scales, due to the mechanism's non-linear nature. Separate universe arguments, however, mean no significant large scale isocurvature will arise on super-horizon scales. Furthermore, any primordial isocurvature will be suppressed due to the thermal equilibrium we assume between the DM and radiation bath~\cite{Weinberg:2004kr,Weinberg:2004kf}.

If there is considerable energy density stored in the gravitational perturbations, reheating effects from small scale Silk damping (diffusion damping)~\cite{Silk:1967kq,Chluba:2012gq,Chluba:2012we} can lead to significant entropy production and dilution of frozen-out relics~\cite{Jeong:2014gna,Nakama:2014vla}. In this paper, we consider a monochromatic perturbation for simplicity, and in that case entropy production can be safely neglected as we shall show below.

The paper is organized as follows. In Sec.~\ref{sec:perturbations}, we introduce our notation for the perturbations and summarize their evolution. In Sec.~\ref{sec:Boltzmann}, we sketch the derivation of the Boltzmann equation governing the acoustically driven DM freeze-out. In Sec.~\ref{sec:results}, we solve this equation numerically, showing the effect of the perturbations on the required cross section, which depends on the amplitude and length scale of the perturbation as well as the partial wave dominating the DM annihilation. In Sec.~\ref{sec:entropy} we discuss the effect of entropy production. We then conclude in Sec.~\ref{sec:conclusion}.

\begin{figure}[t]
\begin{center}
\includegraphics[width=0.45 \textwidth]{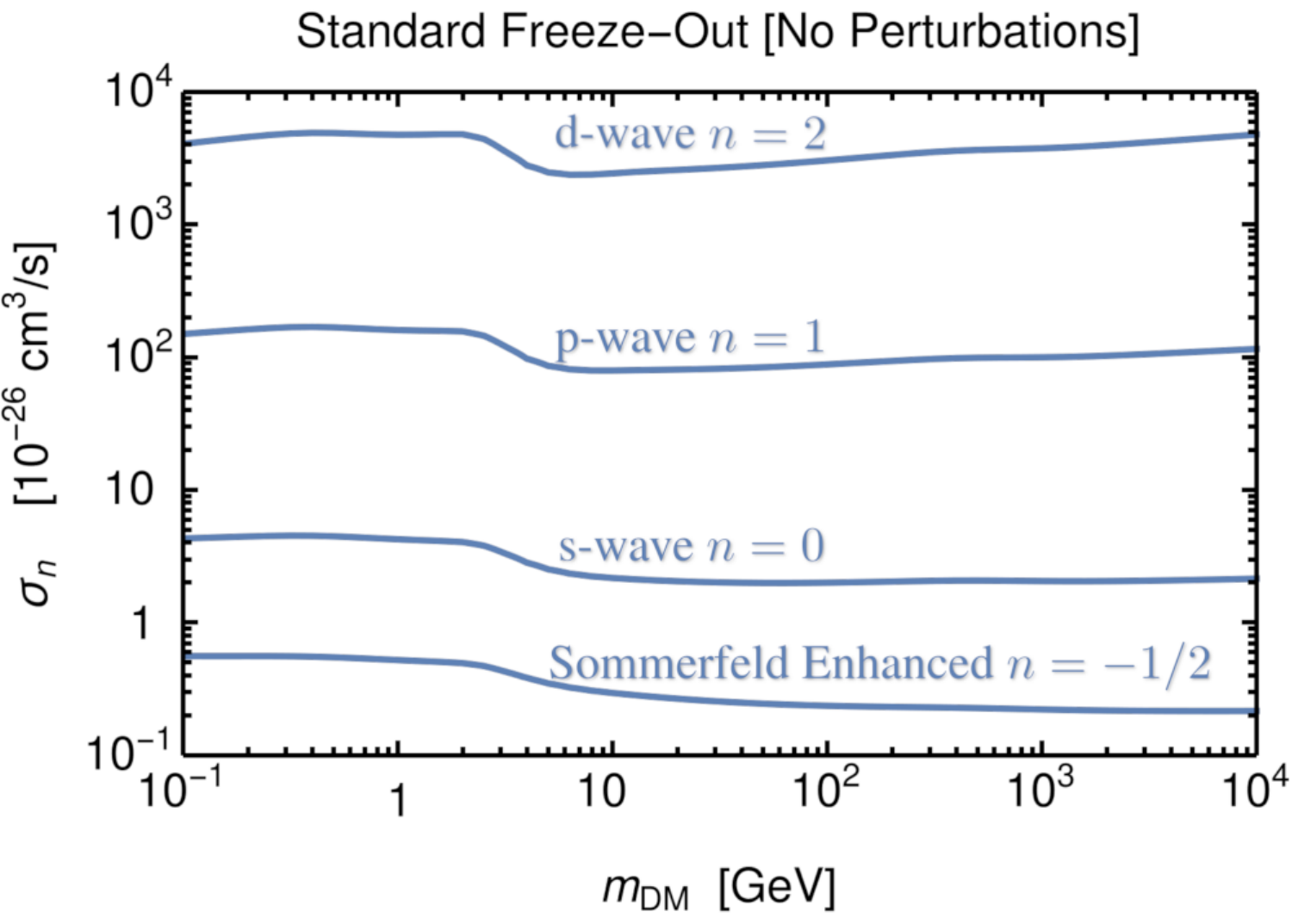} 
\end{center}
\caption{
Required cross section to match the relic abundance in the standard freeze-out scenario with $\langle \sigma v_{\rm rel} \rangle \equiv \sigma_{n} \times ( T/\mdm)^n$ and $\sigma_{n}$ constant. Our $\sigma_0$ values agree closely with Ref.~\cite{Steigman:2012nb}, up to slight differences coming largely from the requirements for $\Omega_{\rm DM} h^{2}$ today. Our results with the perturbations will be normalized to these standard cross sections.}
\label{fig:Standard_FO}
\end{figure}

\section{Perturbations}
\label{sec:perturbations}

Throughout the present study we assume a radiation dominated universe. We define	
	\begin{equation}
	\bar{x} = \frac{ \mdm }{ \bar{T} },
	\end{equation}
where $\mdm$ is the DM mass, and $\bar{T}$ is the averaged background temperature. The gravitational scalar potentials are equal in the absence of anisotropic stress $\Psi = \Phi$, and evolve as~\cite{Peter:2013avv}
	\begin{equation}
	\Psi(\bar{x}) = 2 | \mathcal{R}_{i} | \cos{\delta} \, \frac{ \sin \varphi - \varphi \cos \varphi }{ \varphi^3 }.
	\label{eq:Psi}
	\end{equation}  
Here $\delta$ is a space-coordinate dependent phase, explained in Ref.~\cite{Hotokezaka:2025ewq}, $\mathcal{R}_{i}$ is the amplitude of the primordial curvature perturbation, and 
	\begin{equation}
	\varphi \equiv  \frac{ k \eta }{ \sqrt{3} } \simeq \frac{ \bar{x} }{ \bar{x}_H },
	\label{eq:varphi}
	\end{equation}
where $\eta$ is the conformal time, $k$ is the wavenumber of the oscillation, and $\bar{x}_H$ specifies the time of Horizon entry for the given wavenumber. The approximation relating $\eta$ and $\bar{x}$ above becomes exact in the absence of changing entropic degrees-of-freedom. The total radiation density is $\rho_{r} = \overline{\rho}_{r} + \delta \rho_{r}$, where $\overline{\rho}_{r}$ is the background value, and the fluctuation in the Newtonian gauge is given by 
	\begin{subequations}
	\begin{align}
	\delta_{r}^{N} \equiv \frac{ \delta \rho_{r} }{ \overline{\rho}_{r} } &   = -2 \varphi^{2} \Psi - 2 \eta \Psi' - 2\Psi \\
								& =	-2 \varphi^{2} \Psi - 2 \varphi  \frac{ d \Psi }{ d \varphi } - 2\Psi,
	\end{align}
	\end{subequations}
where prime denotes a derivative with respect to conformal time. (In the co-moving gauge $\delta_{r}^{C} = -2 \varphi^{2} \Psi$.) The corresponding temperature is given by $T = \overline{T} + \delta T$, with fluctuation in the Newtonian gauge~\cite{Peter:2013avv}
	\begin{equation}
	\delta_{T} \equiv \frac{ \delta T }{ \overline{T} }  = \frac{ \delta_{r}^{N} }{ 4 + d \log g_{\ast} / d \log T  }, 
	\label{eq:Tevo}
	\end{equation}
where $g_{\ast}$ are the radiation degrees-of-freedom. It is also useful to define
	\begin{equation}
	x_{T} \equiv  \frac{ \mdm }{ T } =  \frac{ \bar{x} }{ 1 + \delta_T },
	\end{equation} 
for later use. Note fluctuations with $|\mathcal{R}_{i}| \lesssim 0.33$ correspond to $\delta_{r}^{C} \lesssim 0.4$ at Horizon crossing. The threshold for primordial black hole formation is typically $0.4 \lesssim \delta^{C}_{r} \lesssim 0.66$~\cite{Musco:2012au,Harada:2013epa}. We limit ourselves to $|\mathcal{R}_{i}| \lesssim 0.3$ in this paper, so that primordial black hole formation is not an issue. Finally it is also useful to consider the velocity perturbation, $\delta U^{\mu} \equiv a^{-1}(-\Psi,v^{i})$ where $a$ is the scale factor, and we are working in the Newtonian gauge. In this gauge we have, $v_{i} = \partial_i V$ at first order in perturbation theory and assuming zero fluid vorticity, where the evolution of the scalar velocity potential is given by~\cite{Peter:2013avv}
	\begin{equation}
	kV = -\frac{ \sqrt{3} }{ 2 } \varphi \left( \varphi \frac{ d \Psi }{ d \varphi } + \Psi \right).
	\label{eq:kV}
	\end{equation}
If we consider an observer at rest with the fluid at the background level, then the perturbation $v_i$ will also be equivalent to the bulk fluid three-velocity.

\begin{figure*}[t]
\begin{center}
\includegraphics[width=0.32 \textwidth]{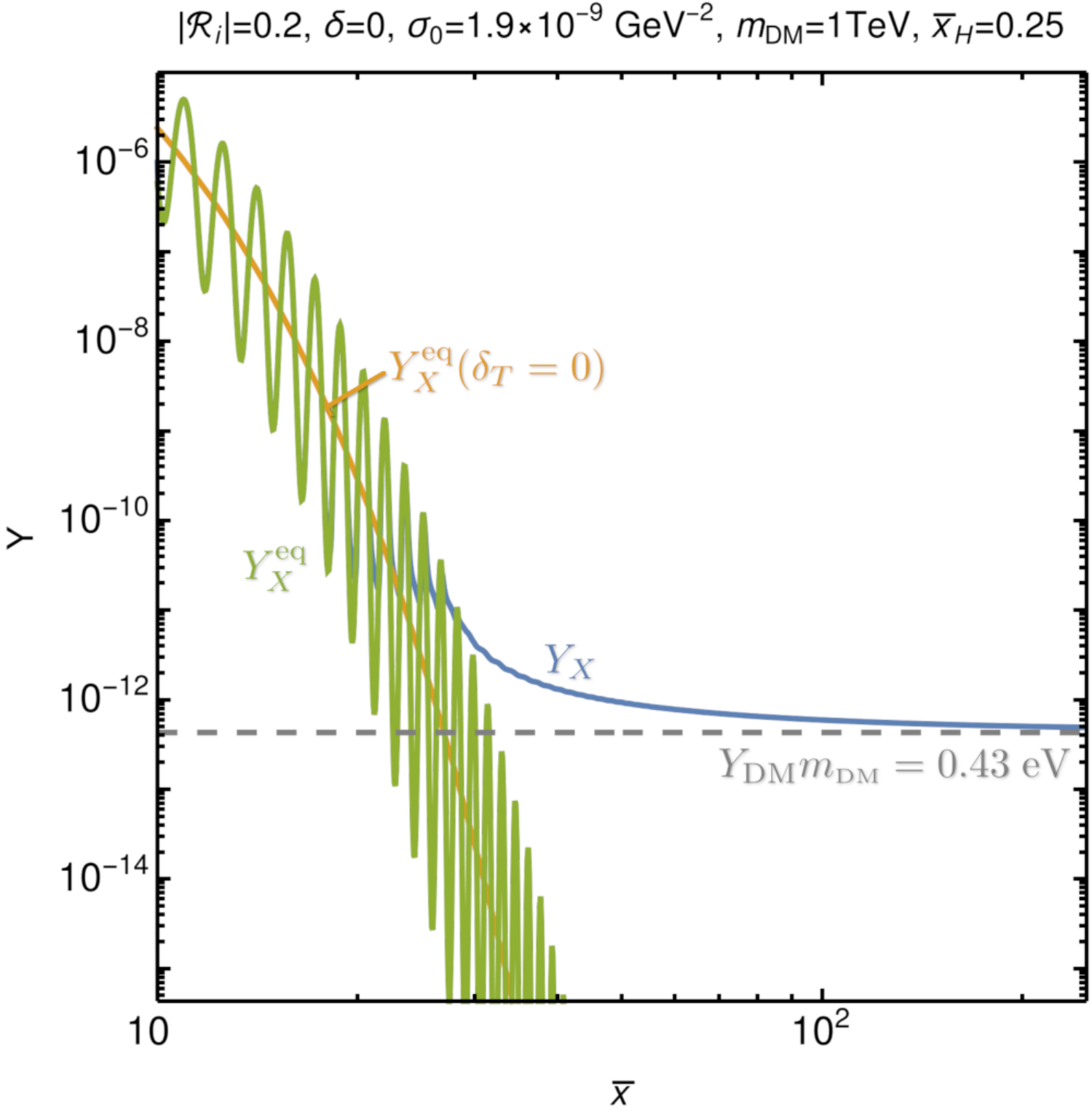} \,
\includegraphics[width=0.32 \textwidth]{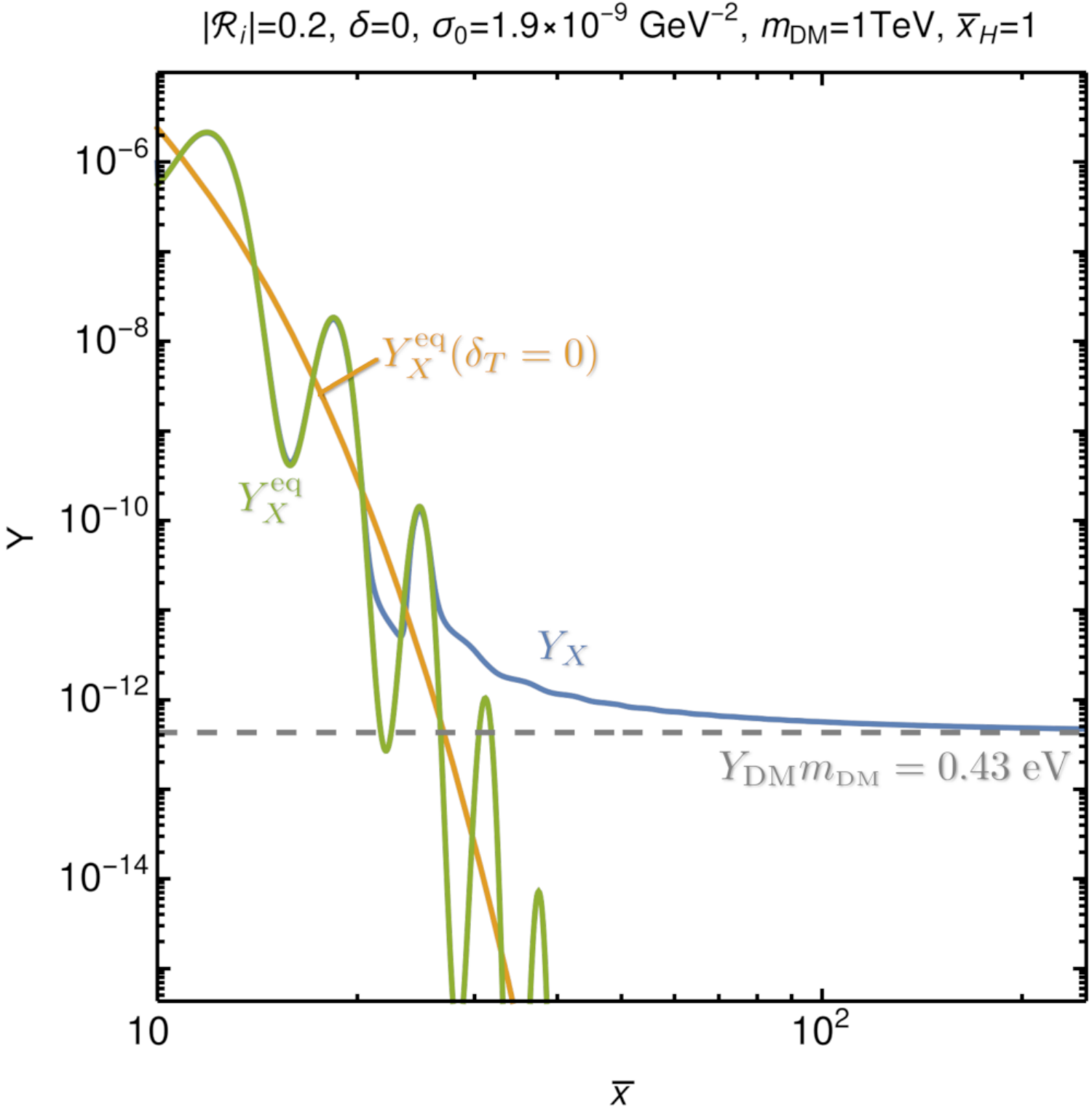} \,
\includegraphics[width=0.32 \textwidth]{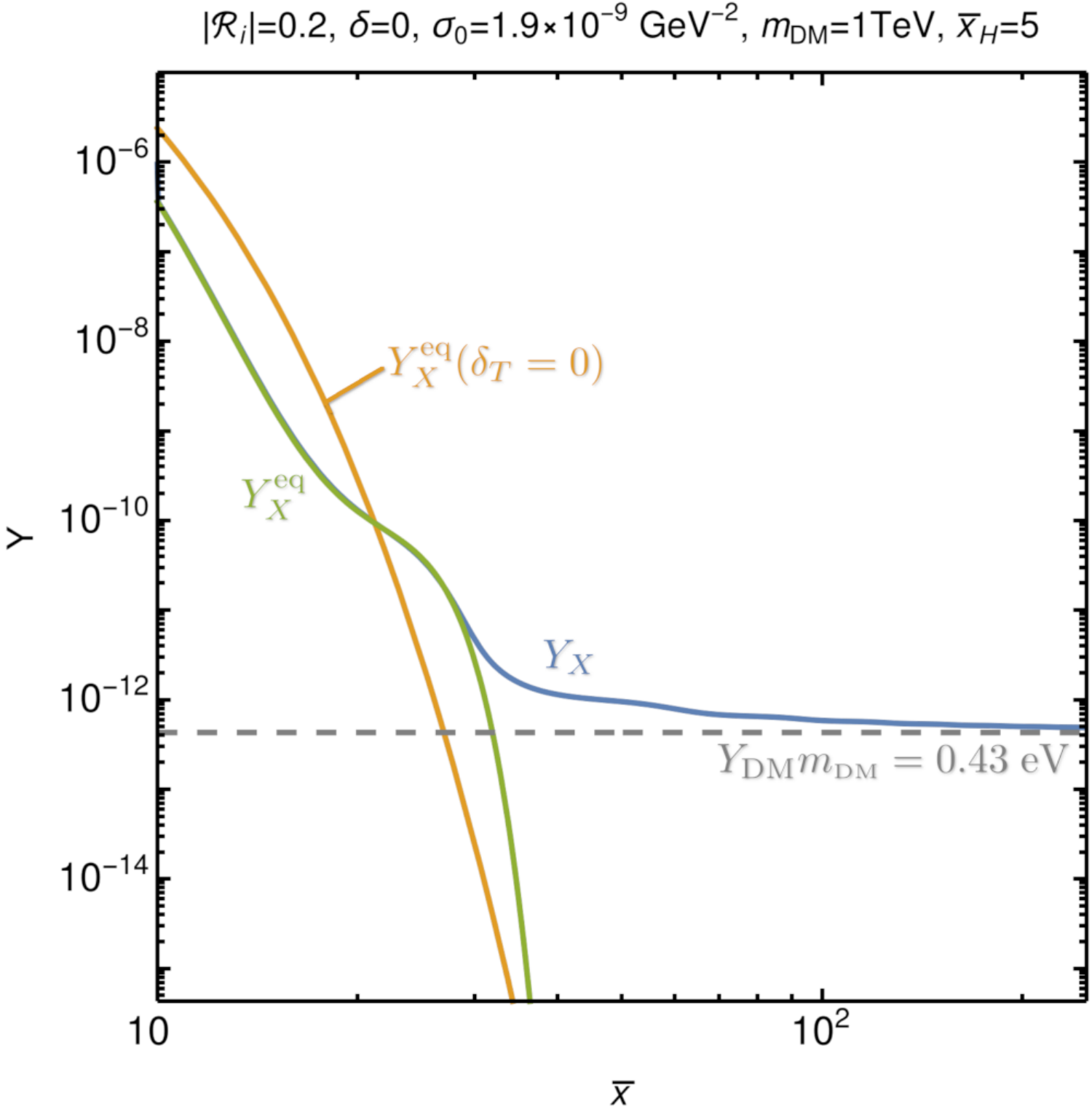}
\end{center}
\caption{
Example solutions to the Boltzmann equations for acoustically driven DM freeze-out with s-wave annihilations.
}
\label{fig:bmannexamples}
\end{figure*}	

\section{DM Boltzmann equation}
\label{sec:Boltzmann}

We assume the DM is in kinetic equilibrium with the radiation bath. We normalize the DM number density to the entropy density
	\begin{equation}
	Y_{X} \equiv \frac{ n_{X} }{ s(T) }.
	\end{equation}
The equilibrium value is given by $Y_{X}^{\rm eq}= n_{X}^{\rm eq}/s$, where the equilibrium number density is given by
	\begin{equation}
	n_{X}^{\rm eq} = \frac{ g_{X} \mdm^2 T  }{ 2 \pi^2  } K_{2}(x_{T}),
	\end{equation}
$K_{2}$ is the modified Bessel function of the second kind, and $g_{X}$ are the DM degrees-of-freedom (we set $g_{X} = 1$ for the purposes of this paper). 
The entropy density is given by
	\begin{equation}
	s = \frac{2 \pi^{2} }{ 45 } g_{\ast S}  T^{3},
	\end{equation}
where $g_{\ast S}$ are the effective entropic degrees-of-freedom (determinations of which can be found, e.g.~in Ref.~\cite{Drees:2015exa}).
Freeze-out is governed by the competing effects of DM annihilations, inverse annihilations, and the Hubble rate. The latter is given by
	\begin{equation}
	H(\overline{T}) = \sqrt{ \frac{ 8 \pi^{3} g_{\ast} }{ 90 } } \frac{ \overline{T}^{2} }{ M_{\rm Pl} },
	\end{equation}
where $M_{\rm Pl} \simeq 1.2 \times 10^{19}$ GeV is the Planck mass. 

Let us now introduce the DM number density current, $n_{X} U^{\mu}$, where $U^{\mu}$ is the four-velocity. Taking the covariant derivative, the evolution of the DM number density with respect to conformal time can be found, via~\cite{Hotokezaka:2025ewq,Senatore:2008vi}
	\begin{align}
	(n_{X} U^{\mu})_{; \mu} & = (1 - \Psi ) \left( \frac{n'}{a} + 3 \mathcal{H} \frac{n}{a} \right) + \frac{n}{a} (v^{i}_{,i} - 3 \Psi')  \nonumber \\
	& = \langle \sigma v_{\rm rel} \rangle (n_{X}^{\mathrm{eq} \, 2}- n_{X}^{2}),
	\label{eq:DMevo}
	\end{align}
where $\mathcal{H} = a'/a = 1/\eta$, and we are working in the Newtonian gauge. In the final line we have equated $(n_{X} U^{\mu})_{; \mu}$ to the integrated collision term. The bulk velocity divergence $v^{i}_{,i}$ appearing in the above equation is the same one as for the radiation, which follows from our assumption of kinetic equilibrium between the two. On the other hand, $v_{\rm rel}$ is the relative velocity of the DM particles coming from their thermal distribution. The local entropy density evolves according to a similar equation,
	\begin{equation}
	(1 - \Psi ) \left( \frac{s'}{a} + 3 \mathcal{H} \frac{s}{a} \right) + \frac{s}{a} (v^{i}_{,i}  - 3 \Psi') = 0,
	\label{eq:entropyevo}
	\end{equation}
assuming conservation of total background entropy (otherwise there would be a non-zero collision or source term on the right hand side). The above equations can be understood in an intuitive way. The $(1 - \Psi )$ pre-factor is a time dilation effect, the $\mathcal{H}$ term captures the dilution of the number density due to the Hubble expansion, the velocity divergence $v^{i}_{,i} = -k^2 V$ takes into account the bulk motion of the fluid leading to changes in the local number density, and the $\Psi'$ term arises from length contractions in the spatial metric components.  Note the fluctuations in the entropy density due to the oscillating energy density, captured in the above equation through the $\Psi, \Psi',$ and $v^{i}_{,i}$ factors, matches the evolution in the entropy density implied by Eq.~\eqref{eq:Tevo} with $s \simeq \bar{s} ( 1 +  3 \delta_T)$, as required by logical consistency. This is checked in App.~\ref{sec:appA}.

To cast the Boltzmann equation into a useful form, tracking $Y_{X} = n_{X}/s$ as a function of $\bar{x}$, one uses the above equations together with
	\begin{equation}
	\frac{d}{d\eta} = a \frac{d}{dt} = a\bar{x} H(\overline{T})  \left( 1 + \frac{1}{3} \frac{ d \log g_{\ast S} }{ d \log \overline{T} } \right)^{-1}  \frac{d}{d\bar{x}} ,
	\end{equation}
which follows from conservation of the averaged entropy (from the above equation one can in principle also determine the corrections from the changing $g_{\ast S}$  to the approximate relation given in Eq.~\eqref{eq:varphi}). The Boltzmann equation governing the DM freeze-out, following the techniques of Refs.~\cite{Hotokezaka:2025ewq,Senatore:2008vi}, is then given by
	\begin{align}
	  \frac{ d Y_{X} } { d \bar{x} } &  = 	\label{eq:DMBmann} \\
	 & \frac{ \langle \sigma v_{\rm rel} \rangle  s(x_{T}) }{ (1-\Psi(\bar{x})) H(\bar{x}) \bar{x} } \left( 1 + \frac{1}{3} \frac{ d \log g_{\ast S} }{ d \log \overline{T} } \right) \left( Y_{X}^{\mathrm{eq} \,  2} - Y_{X}^{2} \right), \nonumber 
	\end{align}
where $Y_{X}^{\rm eq}$ and $\langle \sigma v_{\rm rel} \rangle$ should be evaluated with the local temperature, $T$, i.e.~including the fluctuation.
For the thermally averaged cross section we use the parameterisation
	\begin{equation}
	\langle \sigma v_{\rm rel} \rangle \equiv \frac{ \sigma_{n} }{ x_{T}^{n} } 
	\end{equation}
where $\sigma_n$ is temperature independent and choices $n = -1/2, \, 0, \, 1, \, 2$ correspond to the Sommerfeld-enhanced (SE) Coulomb regime, s-wave, p-wave, and d-wave contact interaction DM annihilations respectively, in terms of which we formulate our results.

\section{Results}
\label{sec:results}

\begin{figure}[t]
\begin{center}
\includegraphics[width=0.43 \textwidth]{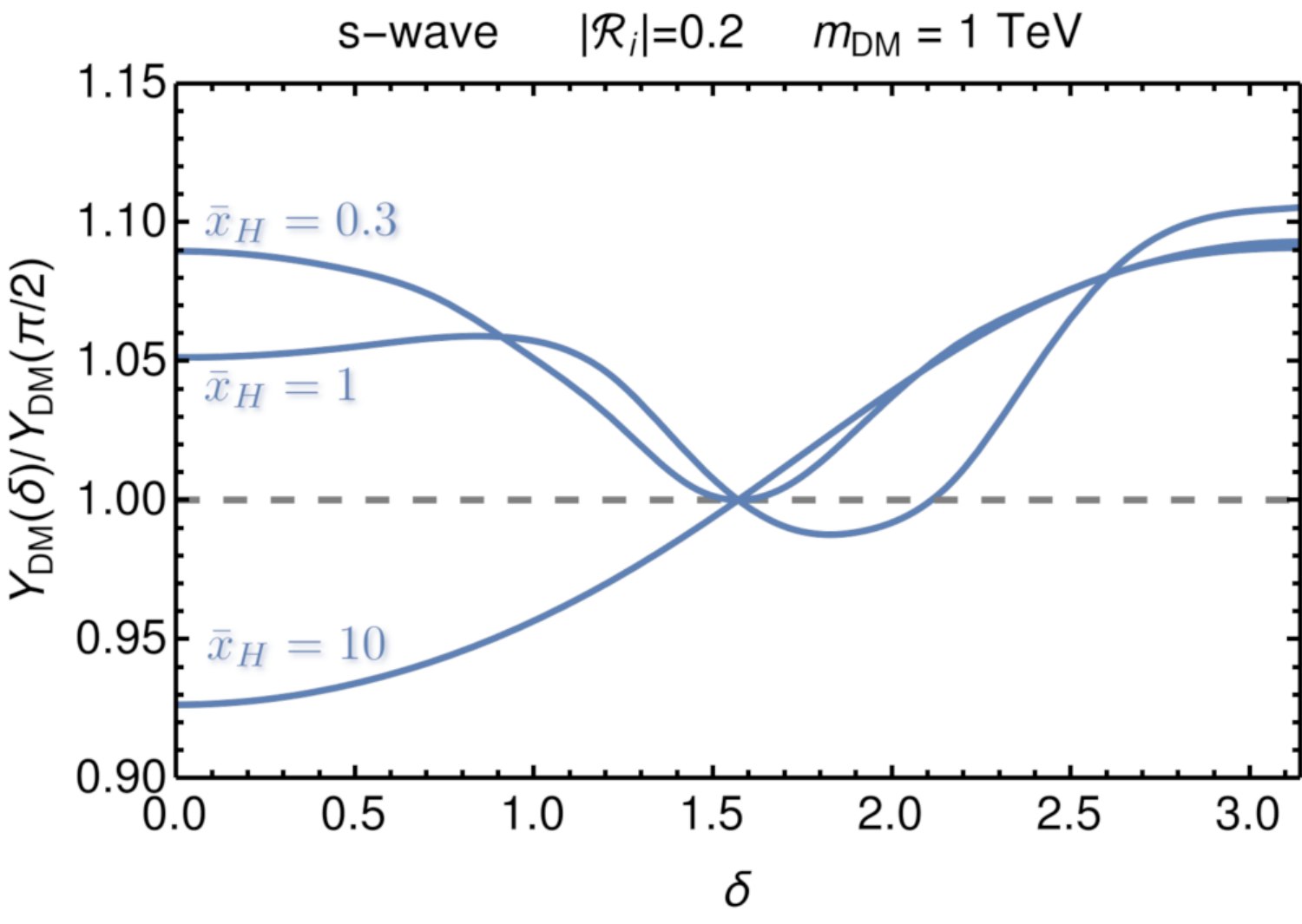} 
\end{center}
\caption{
Example showing dependence of the yield on the phase  $\delta$, which controls the fluctuation amplitude, here assuming s-wave DM annihilations. For smaller wavelengths, illustrated with $\bar{x}_{H}=0.3$ and $\bar{x}_{H}=1$, the yield is enhanced by the presence of the perturbation for most values of $\delta$. In the long wavelength limit, illustrated with $\bar{x}_{H}=10$, there is a cancellation between the yield from under-dense and over-dense regions, which have not had the chance to undergo significant sub-horizon evolution before DM freeze-out.}
\label{fig:deltascan}
\end{figure}

\begin{figure}[t]
\begin{center}
\includegraphics[width=0.45 \textwidth]{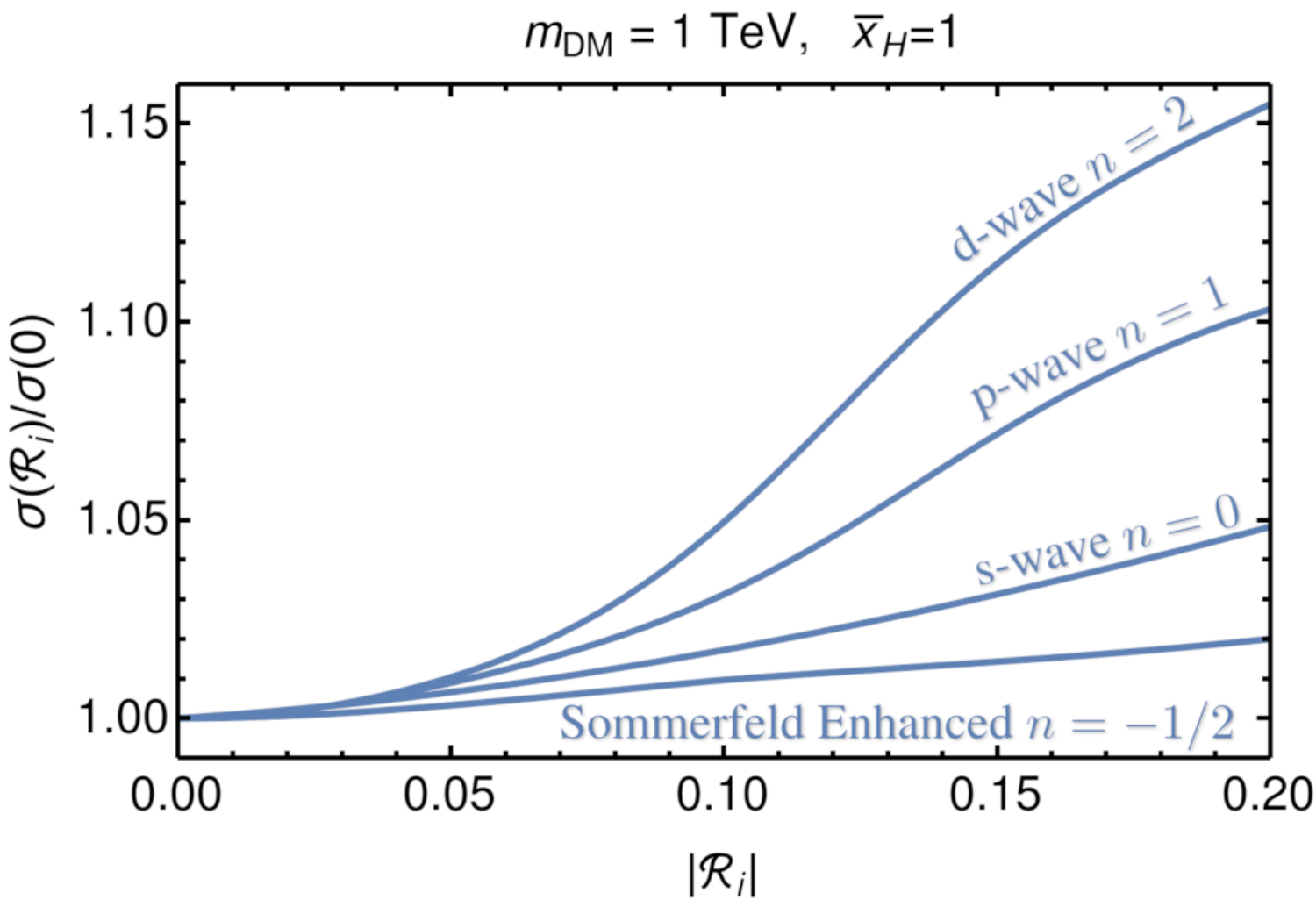} 
\end{center}
\caption{
Dependence of the required cross section, in order to match the observed $Y_{\rm DM} \mdm \simeq 0.43$ eV, as a function of $\mathcal{R}_{i}$. The cross sections decreasing faster with temperature (larger $n$) exhibit a stronger sensitivity to the acoustically driven freeze-out.
}
\label{fig:Ri_plot}
\end{figure}

To aid comparison, we first solve the Boltzmann equation with no perturbations present, and find the required $\sigma_{n}$ to match the observed DM relic abundance. More precisely, we numerically integrate Eq.~\eqref{eq:DMBmann}, starting from equilibrium initial conditions at $\bar{x}=10$, to $\bar{x} = 10^{5}$ at which point $Y_{X}$ is constant (or sufficiently close-to-constant for our purposes), in order to determine the DM evolution. In order to match observations, one requires $Y_{\rm DM} \mdm \simeq 0.43$ eV at late times~\cite{Planck:2018vyg}. For self-conjugate DM one has $Y_{\rm DM} = Y_{X}$, which is the situation we study here. The required cross sections without the perturbations are shown in Fig.~\ref{fig:Standard_FO}. In the non-self conjugate case $Y_{\rm DM} = 2Y_{X}$, and the required cross sections to match the relic abundance increase by a factor of $\simeq 2$. 

We then solve the Boltzmann equation with perturbations. To make the evaluation numerically tractable, we make an approximation to Eq.~\eqref{eq:Tevo} and evaluate $g_{\ast}$ at the average temperature
	\begin{equation}
	\delta_{T} \approx \frac{ \delta_{r}^{N} }{ 4 + d \log g_{\ast} / d \log \overline{ T }   },
	\end{equation}
which will only lead to small errors in parameter space in which $g_{\ast}$ is changing appreciably at the time of DM freeze-out. Some example solutions of the Boltzmann equations with acoustically driven freeze-out are shown in Fig.~\ref{fig:bmannexamples}. The behaviour of $Y_{X}$ can be understood in a simple way. When in chemical equilibrium, the fluctuation in the radiation $T$ increases the energy of the bath particles, allowing more of the high energy tail of the phase space distributions to be accessed and increasing $n_{X}^{\rm eq}$. Thus an increase in $T$ also increases the number of DM particles, when in chemical equilibrium, by reducing the Boltzmann suppression factor. After freeze-out, the DM is chemically decoupled, one can disregard the collision term in the Boltzmann equation, and $Y_{X}$ is constant. Note both $s$ and $n_{X}$ are still fluctuating, as can be seen via inspection of Eqs.~\eqref{eq:DMevo} and \eqref{eq:entropyevo}, but these fluctuations are the same in kinetic equilibrium so that $Y_{X}$ is constant. 

Having the ability to solve the Boltzmann equation, we then average over the phase $\delta$, appearing in Eq.~\eqref{eq:Psi}, which corresponds to taking a spatial average. This allows us to find the overall DM yield. A plot showing examples of the dependence of the yield on $\delta$ is shown in Fig.~\ref{fig:deltascan}. For smaller $\bar{x}_H$ the yield is found to be enhanced by the presence of the perturbation for most values of $\delta$, while for larger $\bar{x}_H$ there is a cancellation as expected between under-dense and over-dense regions.
The former behavior is consistent with the one found in Refs.~\cite{Hotokezaka:2025ewq,Baldes:2025uwz} in the context of leptogenesis, while the latter is expected from the fact that perturbations with $\bar{x}_H \gg 1$ have not had the chance to undergo significant sub-horizon evolution before DM freeze-out.

\begin{figure}[t]
\begin{center}
\includegraphics[width=0.45 \textwidth]{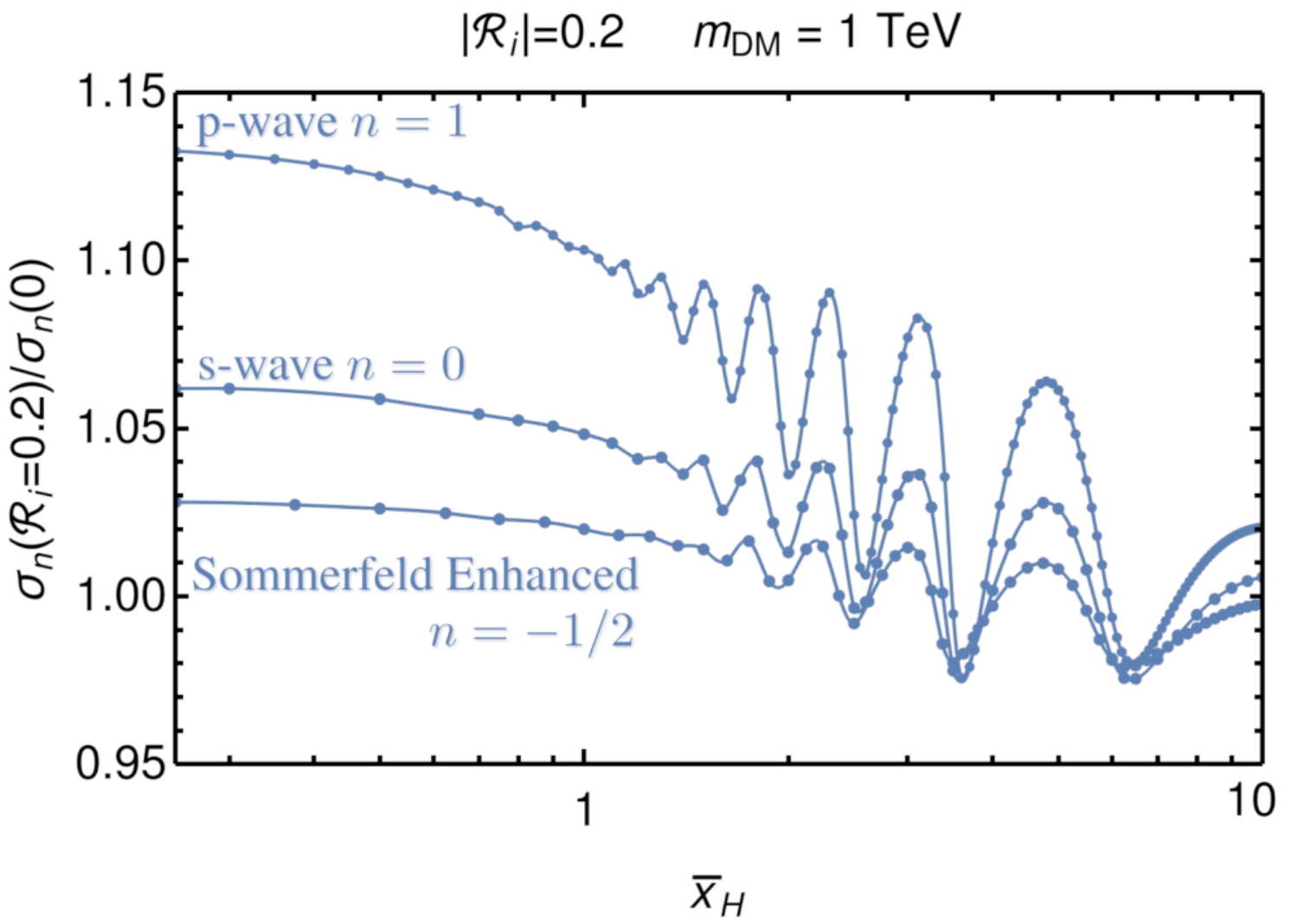} 
\end{center}
\caption{
Dependence of the required cross section, in order to match the observed $Y_{\rm DM} \mdm \simeq 0.43$ eV, as a function of the length scale of the oscillation $x_{H}$. The required cross section is enhanced for small length scales as expected, while for large scales the interplay between the time of freeze-out and the oscillation makes it smaller than the one without perturbations.
}
\label{fig:xH_plot}
\end{figure}

After taking the average over $\delta$, we show the required DM cross section in order to match the observed $Y_{\rm DM} \mdm \simeq 0.43$ eV as a function of $\mathcal{R}_i$ in Fig.~\ref{fig:Ri_plot}.
Since the more abruptly the DM freeze-out occurs the greater the impact of acoustically driven freeze-out, the cross sections decreasing faster with temperature (larger $n$) exhibit a stronger sensitivity.
In Fig.~\ref{fig:xH_plot} we also show the dependence on the length scale of the fluctuation.
For small length scales, the required cross section is enhanced as also expected from Fig.~\ref{fig:Ri_plot}.
On the other hand, for large length scales (but still subhorizon at freeze-out), the interplay between the time of freeze-out and the oscillation means that at some points the required cross section is smaller than the one without perturbations.

We have also checked whether there is any significant dependence on $\mdm$ in the results. As can be guessed already, from the fact that the required $\langle \sigma v_{\rm rel} \rangle$ for DM freeze-out is largely independent of $\mdm$ (up to some $g_{\ast}$ effects) in the standard analysis, the effect of the perturbations is also largely independent of $\mdm$. That is, although we show the plots with the choice $\mdm = 1$ TeV, analogous plots for other choices of $\mdm$ are very similar.

\section{Entropy production}
\label{sec:entropy}

The small scale perturbations are eventually dissipated by diffusion damping, when the length scale is well inside the horizon, and after DM freeze-out. This leads to entropy production at second order in $\delta_{T}$. This has been used to constrain small scale perturbations from the CMB spectral distortions induced by the injection of photons~\cite{Chluba:2012gq,Chluba:2012we}. At even smaller length scales, of interest here, Refs.~\cite{Jeong:2014gna,Nakama:2014vla} find significant effects for BBN, baryogenesis, and DM dilution (albeit with broader, possibly scale invariant spectra for the density perturbations at large $k$). We therefore examine this effect in some more detail. 

To describe the effect, we recapitulate the argument given in Ref.~\cite{Chluba:2012gq}. Consider one Hubble patch with temperature $T_{1} = \bar{T} + \delta T$, and another with temperature $T_{2} = \bar{T} - \delta T$, and ignore the oscillations for now. Once the two patches become enclosed in a single Hubble patch, the average temperature is $\bar{T}$. The average energy density, however, is
	\begin{equation}
	\rho_{\rm avg}  = \frac{ g_{\ast} \pi^{2} }{ 30 } \frac{ (T_{1}^{4} + T_{2}^{4} ) }{ 2 } \simeq \frac{ g_{\ast} \pi^{2} \bar{T}^4 }{ 30 } ( 1 + 6 \, \delta_{T}^2 )  ,
	\label{eq:rhoavg}
	\end{equation}
at second order in the perturbation. Similarly, the average entropy density is
	\begin{equation}
	s_{\rm avg} = \frac{ 2 g_{\ast S} \pi^{2} }{ 45 } \frac{ (T_{1}^{3} + T_{2}^{3} ) }{ 2 } \simeq \frac{ 2 g_{\ast S} \pi^{2} \bar{T}^3 }{ 45 } ( 1 + 3 \, \delta_{T}^2 ) 
	\label{eq:savg}
	\end{equation}
at second order in the perturbation. A black body spectrum with the above entropy would correspond to a temperature
	\begin{equation}
	T_{\rm BB} = \left( \frac{ 45 s_{\rm avg}  }{  2 g_{\ast S} \pi^{2} } \right)^{1/3} \simeq \bar{T} ( 1 + \delta_{T}^2 ).
	\end{equation}
And an energy density
	\begin{equation}
	\rho_{\rm BB} =  \frac{ g_{\ast} \pi^{2} }{ 30 } T_{\rm BB}^{4} \simeq \frac{ g_{\ast} \pi^{2} \bar{T}^4 }{ 30 } ( 1 + 4 \, \delta_{T}^2 ).
	\end{equation}
As $\rho_{\rm avg} - \rho_{BB} = 2 g_{\ast} \pi^{2} \bar{T}^{4} \delta_{T}^2 /30 > 0$, when the average energy density is diffusion damped --- creating a more perfect black body spectrum --- the temperature of the black body is higher than what is given by $T_{\rm BB}$ above. In fact, the temperature associated with a black body with an energy density $\rho_{\rm avg}$ is given by
	\begin{equation}
	T_{\rho \mathrm{avg} }  = \left( \frac{ 30 \rho_{\rm avg} }{ g_{\ast} \pi^{2} } \right)^{1/4} \simeq \bar{T} \left( 1 + \frac{3}{2} \delta_{T}^{2} \right).
	\end{equation}
In a thermalized state, the above temperature implies an entropy density
	\begin{equation}
	s_{\rho \mathrm{avg}} = \frac{ 2\pi^{2} g_{\ast S} T_{\rho \mathrm{avg} }^{3} }{ 45 } \simeq   \frac{ 2\pi^{2} g_{\ast S} \bar{T}^{3} }{ 45 } \left( 1 + \frac{9}{2} \delta_{T}^{2} \right).
	\label{eq:sRH}
	\end{equation}
Thus, following the diffusion damping, the entropy increases by a factor
	\begin{equation}
	\frac{ s_{\rho \mathrm{avg}} }{ s_{\rm avg} } \simeq \left( 1 + \frac{3}{2} \delta_{T}^{2} \right).
	\end{equation}
In the above, we have considered the mixing of two regions, but in reality there is a m\'elange of regions with different initial amplitudes, and there is also energy associated not only with $\delta_{T}$ but also with the velocity of the fluid. Accounting for the energy in the velocity ensures $\rho_{\rm avg}$ indeed redshifts as radiation, without additional oscillations. (Matching onto alternative notation, $\delta_{T} = \Theta_{0}$, and $kV = 3 \Theta_{1}$, where $\Theta_{0}$ and $\Theta_{1}$ are the monopole and dipole of the Fourier transformed temperature perturbation~\cite{Inomata:2016uip,Hu:1994uz}. Furthermore, in the tight coupling limit, the amplitude of $\Theta_{0}$ is a factor $\sqrt{3}$ larger than the amplitude of $\Theta_{1}$, and $\pi/2$ out of phase.) In the above formulas, we must replace~\cite{Chluba:2012gq}
	\begin{subequations}
	\begin{align}
	\delta_{T}^2 \to \langle \Theta^2 \rangle & = \sum_{l} (2l+1) \langle \Theta_l \rangle = \langle \Theta_{0}^2 \rangle + 3 \langle \Theta_{1}^2 \rangle \\
	 					   &  = \langle \delta_{T}^2 \rangle + \frac{1}{3} \langle (kV)^2 \rangle
	\end{align}
	\end{subequations}
where $\langle ...  \rangle$ denotes the spatial average at any fixed late time, when $\langle \Theta^2 \rangle$ is time independent, and we can ignore moments higher than the dipole in the tight coupling limit~\cite{Dodelson:2003ft}. Taking the $\delta$ phase averaged fluctuation at second order, which corresponds to a spatial average, see Eqs.~\eqref{eq:Psi}  -- \eqref{eq:Tevo}, gives
	\begin{equation}
	\langle \Theta^2 \rangle  = \langle \delta_{T}^2 \rangle + \frac{1}{3} \langle (kV)^2 \rangle  = \frac{ |\mathcal{R}_{i}|^{2} }{ 2 }.
	\end{equation}
We therefore arrive at an entropy boost factor~\cite{Jeong:2014gna,Nakama:2014vla}
	\begin{equation}
	\frac{ s_{\rho \mathrm{avg}} }{ s_{\rm avg} } \simeq \left( 1 + \frac{3}{2} \langle \Theta^2 \rangle \right) = \left( 1 + \frac{3}{4} |\mathcal{R}_{i}|^{2} \right),
	\end{equation}
for our monochromatic perturbations. This reduces the required cross section by the same factor, e.g.~by a factor of $\simeq 3 \%$ for $|\mathcal{R}_{i}|=0.2$. But the effect is below the increase in the required cross section from the acoustic driven freeze-out, except for in the Sommerfeld Enhanced case, and outside of $\bar{x}_{H}$ choices with strong cancellations, see Figs.~\ref{fig:Ri_plot} and \ref{fig:xH_plot}. Realistically, there will be other possibly compensating effects at second order in perturbation theory, such as an increase in the Hubble rate through Eq.~\eqref{eq:rhoavg}.

The reason why we find a small entropy effect, while Refs.~\cite{Jeong:2014gna,Nakama:2014vla} find constraining power for small scale perturbations through entropy production, is simply because of the different primordial spectra we assume. We assume a monochromatic perturbations, while Ref.~\cite{Jeong:2014gna} assumes a scale invariant spectrum of enhanced perturbations above co-moving wave number $k > 10^{4} \; \mathrm{Mpc}^{-1}$, relevant for BBN, DM, and baryognesis, and Ref.~\cite{Nakama:2014vla} assumes as top hat function of enhanced perturbations between $10^{4} \; \mathrm{Mpc}^{-1} < k < 10^{5} \; \mathrm{Mpc}^{-1}$, relevant for BBN. These latter assumptions allow for significant entropy production between the creation of the cosmological relic and today. (Also see Ref.~\cite{Inomata:2016uip}.) 

\section{Discussion and conclusions}
\label{sec:conclusion}

Early universe processes may have resulted in large amplitude primordial density fluctuations at small scales. It has been pointed out that a non-linear effect arising from the Boltzmann suppression factor can then play a role in freeze-out processes. This was first studied in the context of the sterile neutrinos of type-I leptogenesis~\cite{Hotokezaka:2025ewq}, and then also type-II and type-III leptogenesis~\cite{Baldes:2025uwz}, in which the heavy states undergo gauge annihilations due to their SM $SU(2) \times U(1)$ charges. Some related work on BBN can be found in Refs.~\cite{Jeong:2014gna,Nakama:2014vla,Inomata:2016uip}. 

Here we extended the study of freeze-out dynamics in the presence of such perturbations to DM. In a model independent approach, we found that the required cross section to match the relic abundance increases by a factor of $\sim 10 \%$, for density perturbations with amplitude $\mathcal{R}_{i} \sim 0.2$. The increase becomes more significant for higher partial waves. There is a non-trivial dependence on the length scale of the perturbation, with longer wavelength perturbations giving smaller effects (due to cancellations between regions with negative and positive temperature fluctuations at freeze-out).

In the present study, the changes in the cross section are modest compared to the standard case, nevertheless, they may be important in scenarios with such fluctuations if the DM indirect detection signature has a strong sensitivity to the precise DM mass and couplings, e.g.~if it lies close to a resonance. This can occur in models in which DM undergoes annihilation via bound-state formation at low relative velocities~\cite{March-Russell:2008klu}, e.g.~in minimal-DM models such as $SU(2)$ quintuplet DM~\cite{Cirelli:2005uq,Mitridate:2017izz,Harz:2018csl,Baumgart:2023pwn} (although such cross-sections also typically receive significant Sommerfeld corrections during freeze-out, which counteracts the effects of the acoustically driven freeze-out). 

A future extension of this study to consider non-monochromatic density perturbations would also be of interest, taking into account example density spectra coming from models of inflation, preheating, topological defects, or phase transitions. In the current study, we have shown that entropy production effects, from the diffusion damping of the oscillations, is subdominant to the increase required in the cross section due to the acoustically driven freeze-out effect. The former effect grows increasingly significant for broader spectra of primordial perturbations~\cite{Jeong:2014gna,Nakama:2014vla}. In the case of BBN there can be additional competing effects~\cite{Inomata:2016uip}. It would be interesting to examine the effect of broader spectra on the acoustically driven freeze-out, as well as to investigate its possible application to super-heavy DM~\cite{BaldesDMfollowup}.

\subsection*{Acknowledgements}
IB was supported by the European Union's Horizon 2020 research and innovation programme under Grant Agreement No.~101002846, ERC CoG ``CosmoChart.''
The work of RJ is supported by JSPS KAKENHI Grant Numbers 23K17687 and 24K07013.

\appendix

\section{Self-consistency of the entropy perturbation}
\label{sec:appA}

In Eq.~\eqref{eq:entropyevo}, we gave the Boltzmann equation for the entropy. For self-consistency, this should match the analytic solution for the same quantity coming from the linearized Einstein equations and fluid conservation equations, found via $s = \bar{s}(1+ 3 \delta_T)$, where the temperature perturbation, $\delta_{T}$, is given in Eq.~\eqref{eq:Tevo}. (The derivation of said equations and their solution in the radiation dominated epoch can, e.g.~be found in Ref.~\cite{Peter:2013avv}). The purpose of this appendix is to explicitly verify the equivalence, as a basic check. For simplicity, we ignore changes in $g_{\ast S}$ in this appendix.

\begin{figure}[t]
\begin{center}
\includegraphics[width=0.45 \textwidth]{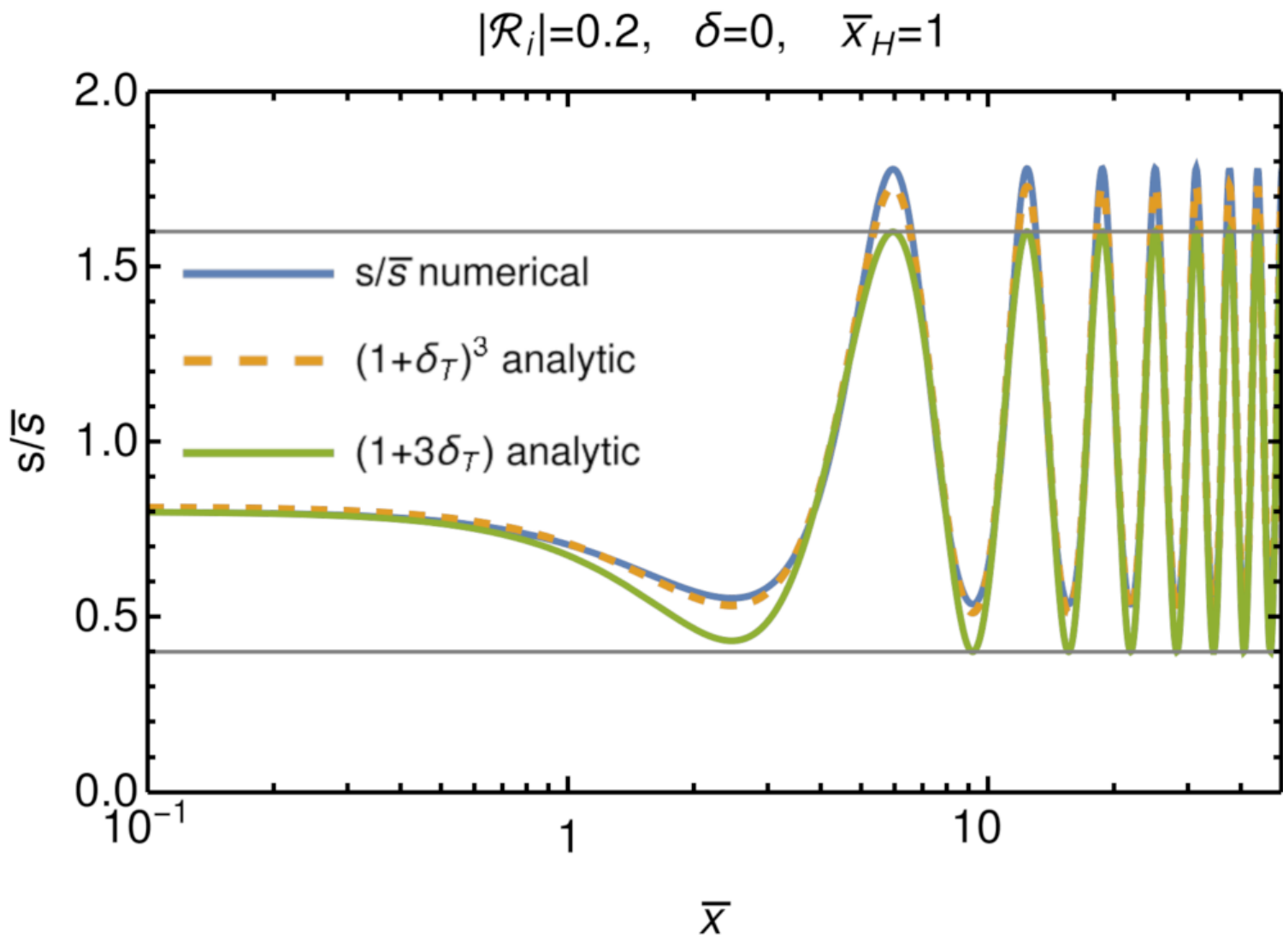} 
\end{center}
\caption{
Comparison of numerical evaluation of the Boltzmann equation for the entropy to the analytic solution. Note the numerical evaluation uses the analytic forms for $kV$ and $\Psi$ as inputs. The comparison is made to show consistency (up to small differences, due to non-linearities at large $|\mathcal{R}_{i}|$). Similar conclusions are also drawn through the purely analytic comparison.}
\label{fig:bmannentropy}
\end{figure}

We start with the Boltzmann equation, Eq.~\eqref{eq:entropyevo}, and re-write it to first order in the perturbations, using $v_{,i}^{i} = -k^2 V$, as
	\begin{equation}
	s' = s ( -3 \mathcal{H} + k^{2}V + 3 \Psi').
	\end{equation}
Now we use $\mathcal{H} = 1/\eta$, $\eta = \sqrt{3} \phi/k$, and $\eta d\varphi  = \varphi d\eta$, to find
	\begin{equation}
	\frac{1}{s} \frac{ ds }{ d \varphi } = -\frac{3}{\varphi} + \sqrt{3} kV + 3 \frac{ d \Psi }{ d\varphi }.
	\end{equation} 
We then substitute in the analytic solutions for $\Psi$ and $kV$, given in Eqs.~\eqref{eq:Psi} and \eqref{eq:kV}, to find
	\begin{align}
	\frac{1}{s} \frac{ ds }{ d \varphi } & = -\frac{3}{\varphi} - \frac{3}{2} \varphi^{2} \frac{ d \Psi }{ d\varphi } - \frac{3}{2} \varphi \Psi + 3 \frac{ d \Psi }{ d\varphi } \label{eq:entropyevo2} \\
					     & = -\frac{3}{\varphi} - \frac{3 | \mathcal{R}_{i} | \cos{\delta} }{ \varphi^{4}} \bigg( 2 \varphi [-3 + \varphi^2] \cos \varphi \nonumber \\
					     &	\qquad \qquad \qquad \qquad \qquad  +[6-4\varphi^2+\varphi^4] \sin \varphi \bigg).  \nonumber
	\end{align}
The above tells us the implied evolution of $s$, given the Boltzmann equation, and the analytic solutions for $\Psi$ and $kV$. 
We now wish to show that $s = \bar{s}(1+ 3 \delta_T)$, and the analytic solutions for $\Psi$ and $kV$, imply the same evolution as \eqref{eq:entropyevo2}, but without using the Boltzmann equation. (Or in other words, that $s = \bar{s}(1+ 3 \delta_T)$, with the analytic solution for $\delta_T$, is also a solution of the Boltzmann equation.) We begin with
	\begin{equation}
	\frac{d}{d\varphi} \left( \frac{ s }{ \bar{s} } \right) = \frac{1}{\bar{s}}  \frac{d s}{d\varphi} - \frac{s}{\bar{s}^2} \frac{d \bar{s}}{d\varphi}  = \frac{1}{\bar{s}}  \frac{d s}{d\varphi} + \frac{3s}{\varphi \bar{s} },
	\end{equation}
where we have used the standard evolution of the average entropy $\bar{s} \propto 1/a^{3}$, which gives $d\bar{s}/d\varphi =-3\bar{s}/\varphi$. We thus have
	\begin{equation}
	\frac{1}{s} \frac{ ds }{ d \varphi } = - \frac{3}{\varphi} +  \frac{ \bar{s} }{ s }\frac{d}{d\varphi} \left( \frac{ s }{ \bar{s} } \right).
	\end{equation}
Now using the analytic solutions we have
	\begin{equation}
	\frac{ s }{ \bar{s} }  \simeq (1 + 3 \delta_{T} ) 
			     = 1 - \frac{3}{2} \varphi^{2} \Psi - \frac{3}{2} \varphi \frac{d \Psi}{d \varphi }  - \frac{3}{2} \Psi.
	\end{equation}
So that at linear order in the perturbation we have
	\begin{align}
	\frac{1}{s} \frac{ ds }{ d \varphi } & =  - \frac{3}{\varphi} +  \frac{d}{d\varphi} \left( \frac{ s }{ \bar{s} } \right)  \\
	& = - \frac{3}{\varphi} - \frac{3}{2} \left( 2 \varphi \Psi + 2 \frac{d \Psi}{d \varphi } + \varphi^2 \frac{d \Psi}{d \varphi } + \varphi \frac{d^2 \Psi}{d \varphi^2 } \right) \nonumber \\
	& = -\frac{3}{\varphi} - \frac{3 | \mathcal{R}_{i} | \cos{\delta} }{ \varphi^{4}} \bigg( 2 \varphi [-3 + \varphi^2] \cos \varphi \nonumber \\
					     &	\qquad \qquad \qquad \qquad \qquad  +[6-4\varphi^2+\varphi^4] \sin \varphi \bigg).  \nonumber
	\end{align}
Which we recognize is the same as the result coming from the Boltzmann equation, i.e.~the second line of~\eqref{eq:entropyevo2}. Alternatively, we can just numerically integrate the Boltzmann equation (after converting Eq.~\eqref{eq:entropyevo} to the variable $\bar{x}$), and compare to the analytic solution for $s(T)$. This is shown in Fig.~\ref{fig:bmannentropy}. The agreement between the numerical solution and the analytic ones improves for smaller $|\mathcal{R}_{i}|$.

\bibliography{Acoustic_DM_FO}

\end{document}